\documentclass{aa}
\usepackage{txfonts}
\usepackage{epsfig}
\usepackage{subfigure}

\usepackage{times}
\usepackage{natbib}
\usepackage{rotating}
\bibpunct{(}{)}{;}{a}{}{,}

\begin{document}

\title{XMM-{\em Newton} spectral properties of the Ultraluminous IRAS 
Galaxy Mrk 273}

\author{I. Balestra, Th. Boller, L. Gallo, D. Lutz, and S. Hess}

\institute{Max-Planck-Institut f\"ur extraterrestrische Physik, Postfach 1312, 
85741 Garching, Germany}

 \offprints{Italo Balestra\\ \email{balestra@mpe.mpg.de}}

\date{Received 2005/ Accepted 2005}

\authorrunning{I. Balestra et al.}

\abstract{We present a 23 ks XMM-{\em Newton} observation of the Ultraluminous 
Infrared Galaxy (ULIRG) Mrk~273. 
The hard X--ray spectrum can be modeled by a highly absorbed 
($\sim7\times10^{23}$ cm$^{-2}$) power law plus an Fe K$\alpha$ emission line. 
The iron line (detected at more than 99\% c.l.) is broad 
($\sigma=0.26^{+0.37}_{-0.17}$ keV), suggesting possible superposition of a 
neutral iron line at 6.4 keV, and a blend of ionized iron lines from Fe XXV 
and Fe XXVI. Given the relatively short exposure, the three line components 
can not be singularly resolved with high statistical significance: the neutral 
component is detected at $\sim2.5\sigma$ and the Fe XXV line at $\sim2\sigma$ 
c.l., while for the Fe XXVI line we can only estimate an upper limit. 
The broad band spectrum requires, in addition to a highly absorbed power law, 
at least three collisionally ionized plasma components, which may be 
associated with star--forming regions. The temperatures of the three plasmas 
are about 0.3, 0.8 and 6 keV, where the highest of the three is 
sufficient to produce ionized iron emission lines. An alternative 
interpretation for the origin of the soft emission might also be given in 
terms of reflection off some photoionized gas, as has been observed in a 
number of nearby Compton--thick Seyfert~2 galaxies (e.g. NGC~1068, Circinus, 
Mrk~3, NGC~4945). A hot gas, photoionized by the primary, continuum can also 
produce ionized iron lines. 
Unfortunately, given the limited statistics and the lack of high resolution 
spectroscopy, it is not possible to distinguish between the two models 
investigated. We further compare the XMM-{\em Newton} findings with the 
{\em Chandra} data obtaining consistent spectral results. 
The absorption corrected hard X--ray luminosity of Mrk~273 is 
$L_{2-10\,\mathrm{keV}}\sim7\times10^{42}$ erg s$^{-1}$, corresponding to 
$\sim0.2$\% of the far--IR luminosity, similar to typical values found in pure 
starbursts. The thermal contribution to the soft X--ray luminosity is 
approximately $0.2-0.7\times10^{42}$ erg s$^{-1}$, comparable to those found 
in NGC~6240 and other starburst dominated ULIRGs. 
We also analyze the XMM-{\em Newton} spectrum of Mrk~273x, an 
unabsorbed Seyfert 2 galaxy at redshift $z=0.458$, which lies in the field of 
view of Mrk~273.
}

\maketitle

\section{Introduction}

The InfraRed Astronomical Satellite (IRAS) has detected 
a very large number of galaxies in the local universe ($\rm z<0.3$) 
which exhibit extraordinarily high infrared luminosity 
($\rm L_{IR}\ge 10^{12}\,L_{\odot}$ 
for $\rm H_0 = 75\,km\,s^{-1}\,Mpc^{-1}$; see Sanders \& Mirabel 1996 for a
review). The number density of these so called Ultraluminous InfraRed Galaxies 
(ULIRGs) exceeds that of optically selected Seyfert galaxies and QSOs with 
comparable bolometric luminosities \citep{Soi87, San88a, San88b, SanMir96} 
by a factor of $\rm \sim 1.5-2$ \citep{San99}. The bulk of luminosity 
in these sources is infrared (IR) emission from warm dust.
Only two dust-heating mechanisms are capable of producing 
such an extraordinary IR luminosity: one involves the presence of a strong 
starburst region, and the other a dust--enshrouded AGN. However it is 
still not clear what the relative contribution of each component to their 
bolometric luminosity should be.

Spectroscopic surveys of samples of ULIRGs carried out with the Infrared 
Space Observatory (ISO) have revealed that about 80\% of such objects are 
predominantly powered by star formation, but the fraction of AGN--powered 
objects increases with luminosity \citep{Gen98, Lut98}.

X--ray observations are a fundamental tool to probe the highly obscured 
innermost regions of these objects, and therefore potentially unveil the 
physical processes at work. In fact a considerable portion of ULIRGs has 
been found to contain a hard X--ray source, highly absorbed by a molecular 
torus, which indicates the presence of a hidden AGN \citep{Mit95, Bra97, 
Kii97, Vig99, Pta03, Brai03, Fra03}.

Mrk~273 is a well studied ULIRG at a redshift $z = 0.03778$ with a Seyfert~2 
nucleus \citep{Kos78, San88a}. It presents evidences for strong star formation 
\citep{Gol95}, such as strong Polycylic Aromatic Hydrocarbons (PAH) features 
in the near--IR \citep{Gen98} and extended ($\sim370$ pc) radio emission, 
punctuated by a number of compact sources identifiable as supernovae or 
supernova remnants \citep{Car00,Bon05}. Like many ULIRGs, Mrk~273 presents 
clear indication of an ongoing merging process such as a long tidal tail and a 
double nucleus \citep{Kna97, Soi00, Car00}. 

ASCA X--ray observations of Mrk~273 showed evidence for a highly absorbed 
($\sim4\times10^{23}$ cm$^{-2}$) hard component above about 3 keV together 
with soft thermal emission \citep{Tur97, Tur98, Iwa99}. 
These findings were indeed confirmed by {\em Chandra}, which revealed 
a compact hard X--ray nucleus inside a much more extended soft 
halo \citep[hereafter Paper~I]{Xia02}.

In this paper we will discuss the analysis of the XMM-{\em Newton} data 
of Mrk~273 and compare them with the previous {\em Chandra} observation 
(Paper I). The data analysis and details of the XMM-{\em Newton} observation 
are described in Section~2, together with our updated reduction of the 
{\em Chandra} data. In Section~3 the spectral fitting results are presented. 
A comparison with the {\em Chandra} observation is given in Section~4.
Our results are then discussed in Section~5 and a summary of the 
XMM-{\em Newton} results on Mrk~273 is given in Section~6. In 
Appendix~A we discuss the XMM-{\em Newton} spectrum of Mrk~273x, an unabsorbed 
Seyfert~2 observed serendipitously in the field of view of Mrk~273. 

Throughout this paper we use the following cosmological parameters: 
$H_0=70$ km s$^{-1}$ Mpc$^{-1}$, $\Lambda_0=0.7$ and $q_0=0$. Within the 
adopted cosmology $1''$ corresponds to 0.75 kpc at the redshift of Mrk~273. 

\section{Observations and Data Reduction}

\subsection{XMM-Newton}

Mrk~273 was observed with XMM-{\em Newton} \citep{Jan01} for $23$~ks on 
2002 May 07 (revolution 0441). During this time the EPIC--PN 
\citep{Str01} and MOS (MOS1 and MOS2; Turner et al. 2001) cameras, 
as well as the Optical Monitor (OM; Mason et al. 2001) and the Reflection 
Grating Spectrometers (RGS1 and RGS2; den Herder et al. 2001) collected data. 
The EPIC--PN and MOS cameras were operated in full-frame mode and utilized 
the thick filter.

The Observation Data Files were processed to produce calibrated event lists 
using the XMM-{\em Newton} Science Analysis System ({\tt SAS v6.1.0}). 
Unwanted hot, dead, or flickering pixels were removed as were events due to 
electronic noise. Event energies were corrected for charge-transfer losses, 
and EPIC response matrices were generated using the {\tt SAS} tasks 
{\tt ARFGEN} and {\tt RMFGEN}. Light curves were extracted from these event 
lists to search for periods of high background flaring. Background flaring was 
negligible. The total good exposure times selected for the PN and MOS were 18 
and 22~ks, respectively.

The source plus background photons were extracted from a circular region with 
a radius of 50$''$, and the background was selected from an off-source region 
with a radius of 70$''$ and appropriately scaled to the source region. Single 
and double events were selected for the PN detector, and single-quadruple 
events were selected for the MOS. The resulting PHA files were grouped with a 
minimum of 20 counts per bin. Pile-up effects were determined to be negligible.

The RGS were operated in standard Spectro+Q mode. The first-order RGS spectra 
were extracted using the {\tt SAS} task {\tt RGSPROC}, and the response 
matrices were generated using {\tt RGSRMFGEN}. Unfortunately, due to the low 
signal-to-noise the RGS data were not fruitful.

The OM was operated in imaging mode for the entire observation. Fourteen 
images were taken in three filters: 5 in $UVW1$ ($245-320$ nm), 9 in 
$UVM2$ ($205-245$ nm), and 1 in $UVW2$ ($180-225$ nm). The average 
exposure time (800~s) was short for such a faint source and magnitudes could 
not be calculated for all $UVM2$ images. The average apparent magnitude in 
each filter was $UVW1=16.40\pm0.07$, $UVM2=16.51\pm0.08$, and 
$UVW2=16.14\pm0.07$, somewhat larger than the ground--based measured U 
($309-373$ nm) magnitude $m_U=15.1\pm0.2$ for the whole galaxy \citep{Sur00}, 
but not inconsistent when accounting for the different aperture sizes used. 

\subsection{Chandra}

$Chandra$ observed Mrk~273 on April 2000 with the Advanced CCD Imaging 
Spectrometer (ACIS-S3 Back Illuminated chip) for about 47 ks. 
The data relevant to this observation were already published in 
Paper~I, where they were processed following the Standard Data Processing 
available at that time (R4CU5UPD13.2, January 2001).

Here, data are reprocessed using much more recent versions of the 
Chandra Interactive Analysis of Observations software (CIAO 3.2) and 
Chandra Calibration Database (CALDB 3.0.0). The most effective improvements 
introduced concern the more accurate background subtraction and more 
careful computation of the effective areas and response matrices.

We start processing data from the level=1 event file. We apply the 
recently released, time--dependent gain 
correction\footnote{http://asc.harvard.edu/ciao/threads/acistimegain/}, 
which is necessary to adjust the ``effective gains", which have been 
drifting with time due to increasing charge transfer inefficiency. 
Since the observation was taken in the VFAINT mode, 
we run the tool {\tt acis\_process\_events} to flag probable
background events using all the information of the pulse heights in a
$5\!\times \!5$ event island to help distinguishing between good
X--ray events and bad events that are most likely associated with
cosmic rays. With this procedure, the ACIS particle background can be
reduced significantly compared to the standard grade selection\footnote
{http://asc.harvard.edu/cal/Links/Acis/acis/Cal\_prods/vfbkgrnd/}.
Real X--ray photons are practically not affected by such cleaning
(only about 2\% of them are rejected, independently of the energy
band, provided there is no pileup).

The data are filtered to include only the standard event grades 
0, 2, 3, 4 and 6. We finally filter time intervals 
with high background by performing a 3$\sigma$ clipping of the 
background level using the script 
{\tt analyze\_ltcrv}\footnote{http://cxc.harvard.edu
/ciao/threads/filter\_ltcrv/}. This yield an effective
exposure time of about 43 ks for the ACIS--S3 chip in the 
energy range $0.3-10$ keV.

\begin{figure}
\begin{center}
\includegraphics[width=8.5 cm, angle=0]{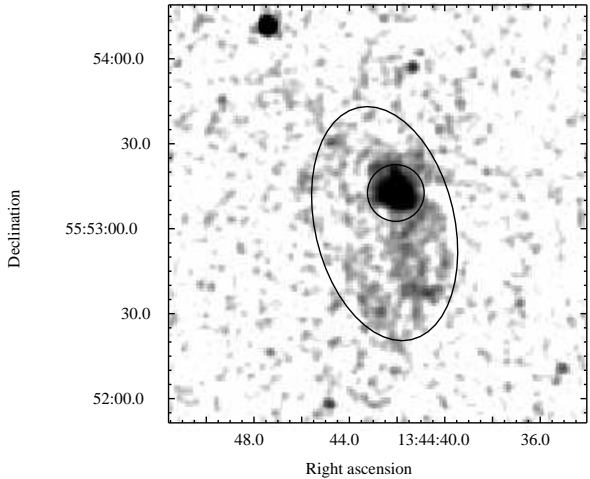}
\caption{Adaptively smoothed $0.3-10$ keV Chandra image of Mrk~273 with a 
smoothing scale of 2 pixels ($\approx1''$). The circle has a radius of $10''$ 
and it is centered at the peak of the hard X--ray emission. The ellipse has a 
size of $42'' \times 25''$ and a position angle of $13^{\circ}$. The bright 
source visible in the upper left corner is Mrk~273x.}
\label{cha_reg}
\end{center}
\end{figure}

As in Paper~I, we separate the spectral analysis of the nuclear emission 
from the extended soft X--ray halo. The spectrum of the nuclear region of the 
source is extracted from a circular region of radius $10''$ centered at the 
peak of the hard X--ray emission, while, for the extended soft halo, we choose 
the region lying between the inner circle and the ellipse shown in 
Figure~\ref{cha_reg} in order to have the best possible signal--to--noise 
ratio.

We use the events included in each of the extraction regions defined 
above to produce a spectrum (PHA) file, which we grouped with a 
minimum of 20 counts per bin, unless otherwise specified. The background is 
obtained from a source free circular region of radius $50''$ located on the 
same chip. 
The background file is then scaled to the source file by the ratio of the 
geometrical area. The response matrices and the ancillary response matrices 
of each spectrum are computed with {\tt acisspec} for the same regions from 
which the spectra are extracted. 

The X--ray spectral analysis has been performed using \textsc{Xspec} 
version 11.3.1. 

The reported errors on model parameters are at the 90\% confidence level for 
one interesting parameter ($\Delta \chi^2=2.71$).

\begin{table}

\caption{XMM-{\em Newton} and {\em Chandra} net number of counts (background 
subtracted) collected by each instrument in the specified extraction region 
and spectral range.}

\begin{center}
   \begin{tabular}{llll}

    \hline\hline 
  
     \bf{Instrument} & \bf{Extraction} & \bf{Energy} & \bf{Net Number} \\
      & \bf{Region} & \bf{[keV]} & \bf{of Counts} \\    
     \hline

    \bf{XMM-Newton} & & & \\ 
    PN & $50''$ & $0.3-10$ & $1680$ \\
    MOS 1 & $50''$ & $0.3-10$ & $590$ \\
    MOS 2 & $50''$ & $0.3-10$ & $520$ \\
      \hline

     \bf{Chandra} & & & \\ 
     ACIS-S & $10''$ & $0.3-10$ & $1990$ \\
     ACIS-S & ellipse & $0.3-2$ & $530$ \\ 
       \hline

\label{counts}
\end{tabular}
\end{center}
\end{table}

\section{Spectral Analysis}

In Table \ref{counts} we give a comparison between the XMM-{\em Newton} 
and {\em Chandra} data sets. The {\em Chandra} ACIS-S spectrum of the nuclear 
region (inner $10''$) of Mrk~273 is dominated by the background above 8 keV, 
while the spectrum of the extended soft X--ray halo is source 
dominated only between 0.5 and 0.9 keV. Both the EPIC--PN and the two MOS 
spectra are source dominated between 0.3 and 10 keV, therefore the entire 
energy range can be analyzed with relatively good signal--to--noise. 
XMM-{\em Newton} therefore provides some further constraints on the shape of 
the continuum between 8 and 10 keV.

The extracted PN and ACIS-S light curves
do not show any significant (within 30\%) flux or spectral variability. 

Throughout our spectral analysis we fix the abundances of the elements to the 
solar values \citep{And89} and we model Galactic absorption with the 
photoelectric absorption model \textsc{PHABS} in \textsc{Xspec}, where we 
fixed the absorbing column density to 
N$\mathrm{_{g}}=1.09\times10^{20}$ cm$^{-2}$ (HEASARC 
W3nH\footnote{http://heasarc.gsfc.nasa.gov/cgi-bin/Tools/w3nh/w3nh.pl}).
Line energies are given in the rest frame of the source. 

\subsection{The complex Fe K$\alpha$ line}

\begin{figure}

\centerline{\epsfig{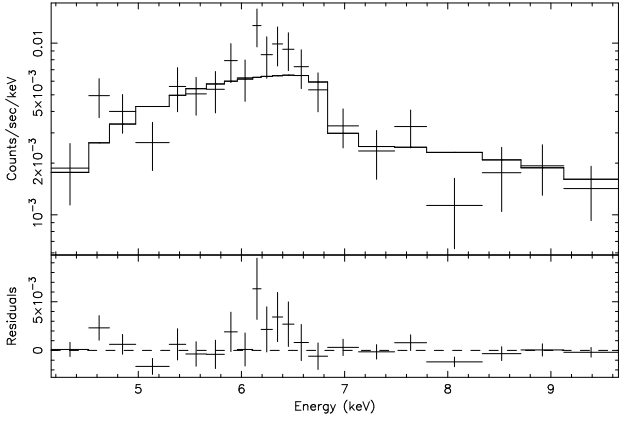}}
\caption{The $4.0-10$ keV EPIC--PN spectrum fitted with a simple absorbed 
power law model ($\Gamma$ is fixed to 1.9). Residuals are found at 
Fe~K$\alpha$ emission line energy (at about 6.5 keV).}
\label{hardpw}

\end{figure}

In order to study the properties of the Fe K$\alpha$ line we concentrate our 
spectral analysis only on the hard X--ray ($4-10$ keV) spectrum of the 
EPIC--PN. In this energy range the continuum can not be reproduced by a simple 
unabsorbed power law, since this model gives a negative spectral index and 
a poor fit. Allowing for extra absorption at the redshift of the source 
(\textsc{ZPHABS} in \textsc{Xspec}), we find a much better fit result 
($\chi^2_r\simeq1.2$), but, given the limited statistics available, if both 
$\Gamma$ and N$\mathrm{_{H}}$ are left free to vary, we find very large 
uncertainties (e.g. $\Gamma=2.4^{+1.2}_{-0.7}$). Fixing the spectral index to 
the ``canonical'' value $\Gamma=1.9$ allows us to measure a column density 
N$\mathrm{_{H}}=(6.2\pm1.5)\times10^{23}$ cm$^{-2}$ intrinsic to the source. 
As displayed in Figure~\ref{hardpw}, the fit is still not completely 
satisfactory ($\chi^2/dof=24.5/20$), mainly because of the presence of a 
feature in the spectrum at about 6.4 keV, which can be identified with an 
Fe K$\alpha$ emission line. A significant improvement in the fit is obtained 
when a single broad Gaussian is added to the model ($\chi^2/dof=15.2/17$). 
The line centroid is at $6.51\pm0.19$ keV and the line is found to be 
broad ($\sigma=0.26^{+0.37}_{-0.17}$ keV), with an equivalent width 
$EW=560^{+490}_{-330}$ eV.

Such a broad iron line (FWHM$\sim29\,000$ km s$^{-1}$) could, in principle, 
originate from the accretion disc, since the primary radiation is obscured 
only up to about $4-5$ keV and therefore should only partly affect the photons 
at the iron line energy. We try to model the iron line profile with the 
\textsc{DISKLINE} model in \textsc{Xspec} for a Schwarzschild black hole, 
which parameterizes the radial emissivity of the disc as a power law (i.e. 
$\propto r^{-q}$). We fix the line rest energy to 6.4 keV, the emissivity of 
the disc $q$ to -2, the inner and outer radii respectively to 6 and to 1000 
$r\mathrm{_{g}}$ and the inclination angle to $30^{\circ}$. We find an 
equivalent width $EW=520\pm300$ eV and a fit result statistically equivalent 
to the one obtained with a single broad Gaussian ($\chi^2/dof=17.1/19$). 
Unfortunately, the statistics are too poor to derive any physical property of 
the disc. 

The hypothesis of a broad iron line originating from the disc 
can not be ruled out, but the high column density, which is obscuring the AGN 
up to about 4 keV, certainly makes the detection of the redshifted wing of 
a relativistic iron line profile much less feasible.
A more likely explanation for the prominent iron line feature observed  
may be given in terms of a complex profile due to the superposition of, at 
least, two unresolved line components: a neutral iron line at 6.4 keV, 
probably originating from a Compton--thick torus, and an ionized (Fe XXV) iron 
line at 6.7 keV. Indeed, the latter could be produced either by some diffuse 
collisionally ionized plasma at temperature $kT\geq5$ keV, as observed in at 
least another two ULIRGs (namely NGC~6240, Boller et al. 2003; Netzer et al. 
2005 and Arp~220, Iwasawa et al. 2005), or by reflection off some photoionized 
gas surrounding the nucleus, as often seen in Compton--thick Seyfert galaxies 
(e.g. Circinus, Sambruna et al. 2001; NGC~1068, Kinkhabwala et al. 2002; 
NGC~4945, Done et al. 2003; Mrk~3, Bianchi et al. 2005). 

When the line profile is modeled with two narrow ($\sigma=0$ eV) Gaussian 
lines with energies fixed at 6.4 and 6.7 keV the result is statistically 
equivalent to a single broad line model ($\chi^2/dof=15.5/18$). In this case 
we measure an equivalent width $EW=165^{+120}_{-110}$ eV and $EW=120\pm95$ eV 
for the neutral and for the ionized component respectively.
Finally, we find no improvement in the fit for the addition of a third ionized 
iron line at 6.97 keV from Fe XXVI. However upper limits on the flux 
($<0.18\times10^{-13}$ erg cm$^{-2}$ s$^{-1}$) and the equivalent width 
($<85$ eV) of this line can be derived.

In the following we discuss two distinct models capable of reproducing the 
broad band spectrum as well as the possible ionized iron lines: one through 
thermal emission, therefore involving the presence of a hot gas associated 
with a starburst, and the other one through reflection of the primary 
continuum by some optically thin photoionized gas. 

\subsection{The thermal emission model}

\begin{table}

\caption[]{XMM-{\em Newton} EPIC--PN spectral fitting results when the thermal 
emission model is applied (2 APEC: two temperature model; 3 APEC: three 
temperature model). The last column displays the result of the combined 
PN+MOS fit with the 3 APEC model. Fluxes are given in units of 
$10^{-13}$ erg cm$^{-2}$ s$^{-1}$, luminosities in units of 
$10^{43}$ erg s$^{-1}$; both are corrected for absorption. 
$^*$ denotes a fixed parameter.}

\begin{center}
\begin{tabular}{llll}

\hline\hline 
  
& \bf{2 APEC} & \bf{3 APEC} & \bf{3 APEC} \\
& \bf{PN} & \bf{PN} & \bf{PN + MOS} \\
\hline

\bf{Apec(1)} & & & \\ 
$kT$ [keV] & $4.7^{+2.2}_{-1.1}$ & $6.1^{+4.3}_{-2.0}$ & $5.4^{+2.6}_{-1.3}$ \\
$f\,_{0.3 - 10\: \mathrm{keV}}$ & $1.40\pm0.15$ & $1.36\pm0.18$ & $1.43\pm0.14$ \\   
\hline
     
\bf{Apec(2)} & & & \\
$kT$ [keV] & $0.66^{+0.06}_{-0.04}$ & $0.75\pm0.07$ & $0.78\pm0.08$ \\   
$f\,_{0.3 - 10\: \mathrm{keV}}$ & $0.63^{+0.11}_{-0.07}$ & $0.55^{+0.12}_{-0.15}$ & $0.47^{+0.13}_{-0.15}$ \\   
\hline

\bf{Apec(3)} & & & \\ 
$kT$ [keV] & -- & $0.26^{+0.08}_{-0.06}$ & $0.31^{+0.09}_{-0.07}$ \\
$f\,_{0.3 - 10\: \mathrm{keV}}$ & -- & $0.23^{+0.14}_{-0.10}$ & $0.23^{+0.13}_{-0.09}$ \\   
\hline

\bf{Power-law} & & & \\
$N_H$ [$10^{22}$cm$^{-2}$] & $68^{+19}_{-15}$ & $69^{+16}_{-19}$ & $69^{+15}_{-14}$ \\
$\Gamma$ & $1.9^*$ & $1.9^*$ & $1.9^*$ \\  
$f\,_{0.3 - 10\: \mathrm{keV}}$ & $44^{+21}_{-13}$ & $44^{+27}_{-13}$ & $46^{+17}_{-12}$ \\ 
\hline
 
\bf{Fe K$\alpha$} & & & \\
$EW$ [eV] & $210^{+190}_{-140}$ & $213^{+198}_{-143}$ & $130\pm120$ \\     
$f\,_{0.3 - 10\: \mathrm{keV}}$ & $0.48^{+0.43}_{-0.34}$ & $0.49^{+0.46}_{-0.33}$ & $0.31\pm0.29$ \\   
\hline

\bf{Total} & & & \\
$f\,_{0.3 - 10\: \mathrm{keV}}$ & 46.6 & 46.6 & 48.7 \\ 
$L\,_{0.3 - 10\: \mathrm{keV}}$ & 1.54 & 1.54 & 1.66 \\ 
\hline
$\chi^2/$dof & 84.2/80 & 63.8/78 & 130.0/134 \\   
\hline
\label{broadband}
\end{tabular}
\end{center}
\end{table}

At first we concentrate our analysis on the EPIC--PN spectrum. The better 
statistics, especially at high energies, permit more robust constraints of the 
model parameters. Once the best fit for the PN is found, we perform a combined 
fit of the PN and the two MOS, applying the same model, in order to further 
improve the statistics.

\begin{figure}

\centerline{\epsfig{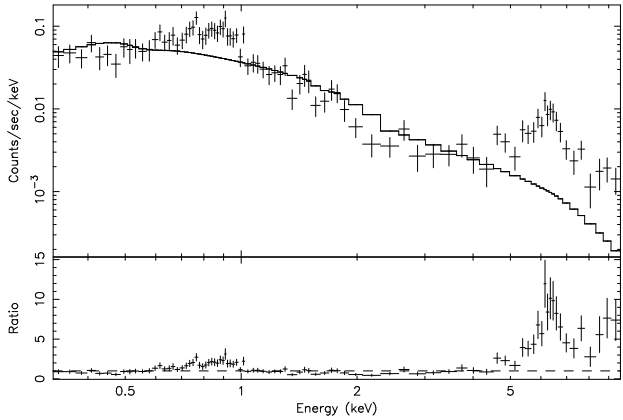}}
\caption{A simple absorbed power law model applied to the EPIC--PN spectrum 
of Mrk 273. Strong residuals (data/model) are clearly present above 4 keV and 
also between 0.6 and 1 keV.}
\label{powlaw_pn}

\end{figure}

When the $0.3-10$ keV EPIC--PN spectrum is modeled with a simple absorbed 
(Galactic plus intrinsic) power law, 
we obtain a statistically unacceptable fit ($\chi^2_r>4$). Strong residuals 
remain primarily above 4 keV, but also between 0.6 and 1 keV (Fig. 
\ref{powlaw_pn}). 

As a starting point for our spectral modelling we adopt the best fit 
absorbed power law plus Gaussian model for the $4-10$ keV band 
(see Section~3.1) and extrapolate to 0.3 keV. A very prominent soft excess is 
clearly noticeable. Since we intend to investigate both the origin of the soft 
excess and of the possible ionized iron, in the following we fix the 
parameters of the Gaussian profile to $E=6.4$ keV and $\sigma=0$ eV. This will 
be considered our baseline model.

Diffuse, optically thin, thermal plasma emission is often observed in 
Starburst galaxies and ULIRGs. It is typically associated with 
star--forming regions and shock--heated gas in merging systems. 
We try to model the soft, extended, X--ray emission with a thermal spectrum 
from a collisionally ionized plasma, using the \textsc{APEC} model 
\citep{Smi01} in \textsc{Xspec}. A single temperature plasma component 
gives a considerable improvement to the fit with respect to the baseline model 
($\chi^2/dof=259/82$), but can not simultaneously account for the observed 
peak at about $0.7-0.9$ keV and the possible ionized iron lines. 

Therefore, we add to the baseline model two collisionally ionized 
plasma components of different temperatures. The free parameters are now the 
temperature and the normalization of each plasma component, the column density 
of the absorber at the redshift of the source, the normalization of the power 
law and the normalization of the Gaussian line. 
The best fit parameters for this model are given in Table~\ref{broadband}. 
A plasma temperature of $\sim0.7$ keV is needed in order to model the peak in 
the emission at low energies, while a much hotter ($\sim5$ keV) plasma is 
required to fit part of the high energy continuum and also to give rise to the 
ionized iron line at 6.7 keV. The second, hotter, plasma is required
with a high statistical significance ($\Delta\chi^2\simeq175$ for the addition 
of 2 free parameters). However the fit is still not completely satisfactory 
($\chi^2/dof=84.2/80$) due to the presence of some excess below about 0.6 keV. 

The addition of a third, lower temperature, plasma component provides a 
further improvement to the fit ($\Delta\chi^2\simeq20$ for the addition of 2 
free parameters). In this case we find an undoubtedly acceptable result 
($\chi^2/dof=63.8/78$). The lowest temperature component is required 
with a probability of more than 99.99\% according to an F--test. Its 
temperature ($kT=0.26^{+0.08}_{-0.06}$ keV) is consistent with the one derived 
from the spectrum of the extended hot gas halo ($kT=0.37\pm0.04$ keV) observed 
in the {\em Chandra} image (see Section~4.2).

\begin{figure}
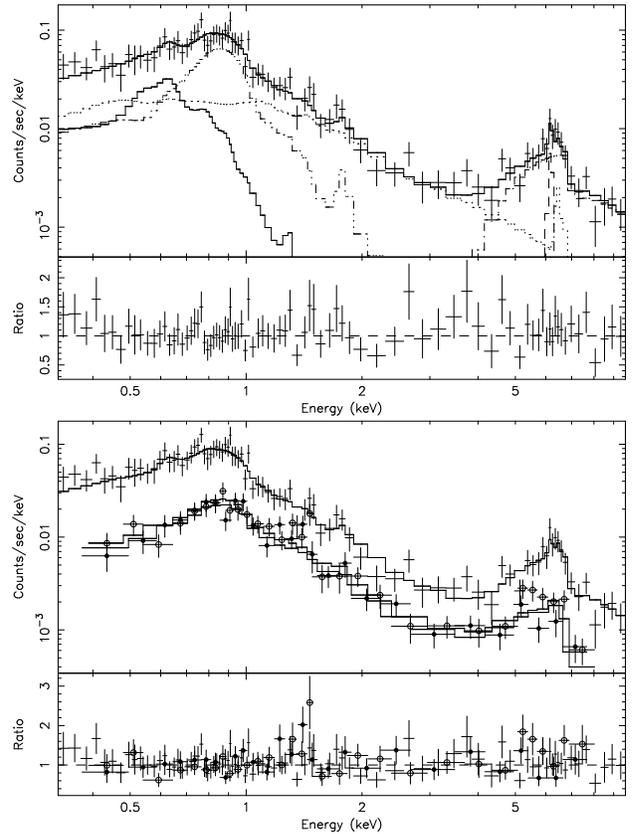


\centerline{\epsfig{figure=xmm_pn_therm.ps,width=5.5cm,angle=270}}
\centerline{\epsfig{figure=model1.ps,width=5.5cm,angle=-90}}
\caption{{\em (Upper Panel)} The EPIC--PN spectrum of Mrk~273 fitted with 
three different temperature APEC models, a highly absorbed power law and a 
neutral Fe K$\alpha$ line. Each model component is displayed. 
{\em (Lower Panel)} The EPIC--PN and the two MOS combined fit with the same 
model. The closed and open circles correspond to the two MOS data sets. 
The best fit parameters are given in Table~\ref{broadband}.}
\label{therm}

\end{figure}

The best fit to the $0.3-10$ keV PN spectrum is therefore obtained using three 
thermal plasma components with different temperatures, a highly absorbed power 
law and a neutral Fe K$\alpha$ line (Figure~\ref{therm}). According to 
this model the emission below about 4 keV would be purely thermal. We will 
discuss the possible implications of these new findings in Section~5. 

Simultaneously fitting the PN and the two MOS data results in an equivalent 
fit, as shown in Figure~\ref{therm}. All the parameters are consistent with 
those obtained using the PN only (see Table~\ref{broadband}). It is worth 
noting that, combining the PN and the MOS, we find a temperature 
$kT=0.31^{+0.09}_{-0.07}$ keV for the coolest thermal component, which is in 
very good agreement with the one measured by {\em Chandra} for the hot 
extended halo.

Finally, we remark that there is no need for further absorption at the 
redshift of the source for any of the three thermal components. When left free 
to vary the absorbing column density of each component is consistent with the 
Galactic value and do not improve the fit ($\Delta\chi^2<1$ for the addition 
of one free parameter). 

\subsection{The photoionized gas model}

An alternative interpretation for the origin of the soft X--ray continuum and 
the ionized iron line observed in the XMM-{\em Newton} spectrum is 
scattering of the primary radiation by some photoionized, Compton--thin, hot 
gas surrounding the AGN. Ionized Fe K$\alpha$ emission lines from 
H-- and He--like iron ions are expected to be produced through recombination 
and resonant scattering \citep{Bia02}. 

The analysis of the {\em Chandra} image in Paper~I revealed that the 
emission below 1 keV is extended on a region of radius $\sim7.5$ kpc 
surrounding the nucleus and, partly, on a much larger region ($\sim45$ kpc) 
encompassing the tidal tail. 
A pure reflection model can not account for such extended X--ray emission, 
which must be of thermal origin. 
If we model the $0.3-10$ keV PN continuum with a highly absorbed (primary) 
plus a less absorbed (reflected) power law having the same spectral index, we 
find an unacceptable fit ($\chi^2_r\simeq2.6$). Therefore we added 
to the model two narrow Gaussian lines, at 6.4 keV from neutral iron and 
at 6.7 keV from Fe~XXV, plus a hot collisionally ionized plasma component. 
In this case the fit result is acceptable ($\chi^2/dof=66.9/79$). However, 
there is no need for an additional lower temperature thermal plasma component, 
which is clearly detected by {\em Chandra}. 

This apparent discrepancy between the two data sets may be reconciled simply 
by taking into account the absorption of the secondary power law, which is 
apparently negligible in the XMM-{\em Newton} spectrum, but 
$\sim7\times10^{20}$ cm$^{-2}$ in the {\em Chandra} spectrum of the 
``nuclear'' region (see Section~4.1). Therefore we fix the column 
density of the material obscuring the reflected power law to the value 
measured by {\em Chandra}. The free parameters of the model are: the 
temperature and the normalization of the thermal component; the column density 
of the absorber at the redshift of the source, which is obscuring only the 
primary power law; the normalizations of the two power law components; and the 
normalizations of the two Gaussian lines. 

When this model is applied to the PN spectrum the fit result is still not 
completely satisfactory ($\chi^2/dof=84.9/80$), mostly due to some excess 
still present below 0.6 keV. The best fit parameters for this model 
are given in Table~\ref{refl}. 

\begin{figure}

\centerline{\epsfig{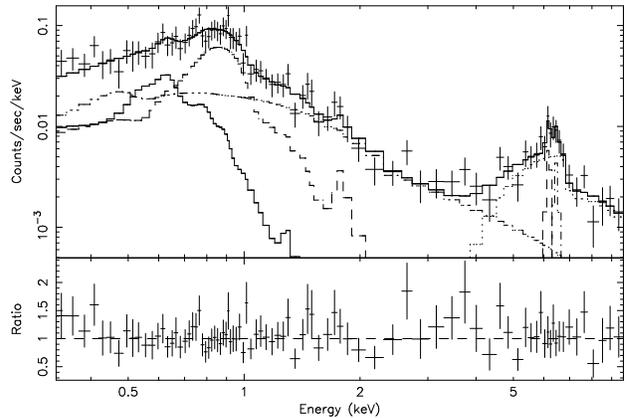}}
\caption{The EPIC--PN spectrum fitted with two thermal plasma components, a 
highly absorbed and a less absorbed power law with the same spectral index, 
plus a narrow Fe K$\alpha$ line at 6.4 keV. The best fit parameters for this 
model are given in Table~\ref{refl}.}
\label{photo}

\end{figure}

The addition of a second, lower temperature, plasma component is a 
considerable improvement to the fit ($\Delta\chi^2\simeq20$ for the addition 
of 2 free parameters, $\chi^2/dof=64.0/78$). Figure~\ref{photo} shows the best 
fit obtained for this model. The cooler thermal plasma 
component is required with a probability of more than 99.99\% according to an 
F--test. The temperature found $kT=0.26^{+0.08}_{-0.03}$ keV is 
consistent, within the uncertainties, with that measured by {\em Chandra} 
for the extended hot gas halo.

The flux of the reflected power law amounts to $2-6$\% of the primary 
component. From this ratio we can estimate the column density of the 
reflecting material. Assuming a covering factor of 0.5, the column density of 
the photoionized gas would be approximately $10^{23}$ cm$^{-2}$. The 
equivalent widths of the ionized iron lines expected to be produced by a hot 
gas with such a density are consistent with the measured values within the 
uncertainties. 

The best fit model obtained here is similar to the one presented in 
Paper~I, with the only noticeable differences being a slightly 
larger absorbing column density obscuring the primary radiation and a 
slightly smaller temperature of the cooler thermal component.

\begin{table}

\caption[]{XMM-{\em Newton} EPIC--PN spectral fitting results using the 
photoionized gas model described in the text. Fluxes are given in units of 
$10^{-13}$ erg cm$^{-2}$ s$^{-1}$, luminosities in units of 
$10^{43}$ erg s$^{-1}$; both are corrected for absorption. 
$^*$ denotes a fixed parameter.}

\begin{center}
\begin{tabular}{llll}

\hline\hline 
  
& \bf{1 APEC} & \bf{2 APEC} \\
\hline

\bf{Apec(1)} & & \\ 
$kT$ [keV] & $0.66^{+0.07}_{-0.04}$ & $0.76^{+0.08}_{-0.07}$ \\
$f\,_{0.3 - 10\: \mathrm{keV}}$ & $0.62\pm0.07$ & $0.52\pm0.11$ \\   
\hline

\bf{Apec(2)} & & \\
$kT$ [keV] & -- & $0.26^{+0.08}_{-0.03}$ \\   
$f\,_{0.3 - 10\: \mathrm{keV}}$ & -- & $0.23^{+0.11}_{-0.10}$ \\   
\hline

\bf{Power-law(1)} & & \\
$N_H$ [$10^{22}$cm$^{-2}$] & $68^{+17}_{-15}$ & $65^{+15}_{-12}$ \\
$\Gamma$ & $1.9^*$ & $1.9^*$ \\  
$f\,_{0.3 - 10 \:\mathrm{keV}}$ & $41_{-12}^{+21}$ & $39_{-11}^{+20}$ \\ 
\hline
 
\bf{Power-law(2)} & & \\
$N_H$ [$10^{20}$cm$^{-2}$] & $7^*$ & $7^*$ \\
$\Gamma$ & $1.9^*$ & $1.9^*$ \\  
$f\,_{0.3 - 10 \:\mathrm{keV}}$ & $1.7\pm0.2$ & $1.5\pm0.2$ \\ 
\hline
 
\bf{Fe K$\alpha$} & & \\
$E$ [keV] & $6.4^*$ & $6.4^*$ \\
$EW$ [eV] & $186_{-125}^{+165}$ & $185_{-122}^{+159}$ \\     
$f\,_{0.3 - 10\: \mathrm{keV}}$ & $0.50^{+0.44}_{-0.34}$ & $0.47^{+0.40}_{-0.31}$ \\   
\hline

\bf{Fe XXV} & & \\
$E$ [keV] & $6.7^*$ & $6.7^*$ \\
$EW$ [eV] & $121_{-102}^{+110}$ & $124_{-103}^{+110}$ \\     
$f\,_{0.3 - 10\: \mathrm{keV}}$ & $0.34^{+0.31}_{-0.29}$ & $0.33^{+0.29}_{-0.27}$ \\   
\hline

\bf{Total} & & \\
$f\,_{0.3 - 10\: \mathrm{keV}}$ & 44.5 & 42.1 \\   
$L\,_{0.3 - 10\: \mathrm{keV}}$ & 1.47 & 1.39 \\ 
\hline
$\chi^2/$dof & 84.9/80 & 64.0/78 \\   
\hline

\label{refl}
\end{tabular}
\end{center}
\end{table}

\section{Comparison with {\em Chandra}}

For a detailed spatial analysis of the {\em Chandra} X--ray image of Mrk~273 
we refer the reader to Paper I. Here, as in Paper I, we extract 
separately the spectra of the nuclear region and that of the extendend soft 
halo from the two regions shown in Figure~\ref{cha_reg}, in order to have the 
best possible signal--to--noise. 

\subsection{The inner 10$''$ region}

We apply to the {\em Chandra} spectrum of the inner region 
($10''\sim7.5$~kpc surrounding the AGN) both models successfully tested in 
Section~3 for the XMM-{\em Newton} data.

\begin{figure}

\centerline{\epsfig{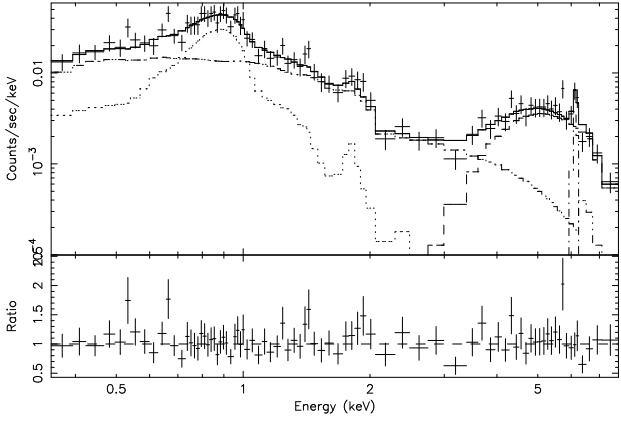}}
\caption{The $0.3-8$ keV ACIS--S spectrum of the inner $10''$ region fitted 
with two thermal plasma components, a highly absorbed power law 
and a narrow Fe K$\alpha$ line at 6.4 keV. The best fit parameters are given 
in Table~\ref{chares}.}
\label{mod1}

\end{figure}

The thermal emission model, which was not tested in Paper~I, 
when applied to the {\em Chandra} spectrum of the ``nuclear'' region, gives 
also a good fit ($\chi^2/dof=65.0/75$). The best fit 
parameters are given in Table~\ref{chares}. In this case, only two 
collisionally ionized plasma components are required, as shown in 
Figure~\ref{mod1}. 

\begin{figure}

\centerline{\epsfig{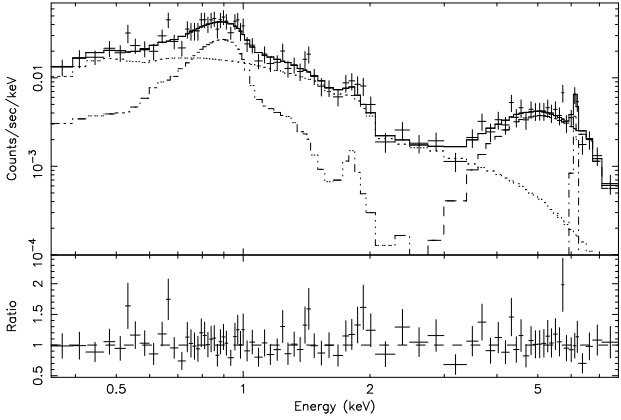}}
\caption{The $0.3-8$ keV ACIS--S spectrum of the inner $10''$region fitted 
with a single thermal plasma component, a highly absorbed and a less 
absorbed power law with the same spectral index, plus a narrow Fe K$\alpha$ 
line at 6.4 keV. The best fit parameters are given in Table~\ref{chares}.}
\label{mod2}

\end{figure}

The photoionized gas model, which we discuss here, is similar to the 
one presented in Paper~I. It consists of a highly absorbed (primary) plus a 
less absorbed (secondary) power law having the same spectral index, two 
Fe K$\alpha$ lines at 6.4 and 6.7 keV and a single thermal plasma component. 
This model also provides a good fit ($\chi^2/dof=62.0/75$, see 
Figure~\ref{mod2}). A column density of 
N$\mathrm{_{H}}=6.8^{+2.3}_{-2.8}\times10^{20}$ cm$^{-2}$ obscuring 
the secondary power law is needed, in agreement with Paper~I and the 
XMM-{\em Newton} results. 

A narrow, unresolved, Fe K$\alpha$ line is detected at about 3$\sigma$ 
significance in the ACIS--S spectrum. Its centroid is found at 
$6.34\pm0.04$ keV, which is slightly lower than the expected value of 6.4 keV 
from neutral iron, probably due to uncertainties in the calibration. 
No ionized iron lines are required by the data; however, we derive upper 
limits on the fluxes of the Fe XXV and Fe XXVI lines (respectively 
$<0.37\times10^{-13}$ and $<0.39\times10^{-13}$ erg cm$^{-2}$ s$^{-1}$), 
which are consistent with those measured by XMM-{\em Newton}.

The best fit parameters for both models (summarized in Table~\ref{chares}) are 
consistent, within the uncertainties, with those found by XMM-{\em Newton}. 
The only exception is in the value of the column density obscuring the primary 
radiation, which is slightly different between the two 
observations ($41\pm6$ and $69^{+16}_{-19}\times10^{22}$ cm$^{-2}$ for the 
{\em Chandra} and the XMM-{\em Newton} observation respectively).

\begin{table}

\caption[]{{\em Chandra} ACIS--S spectral fitting results when the thermal 
emission model (Model~1) and the photoionized gas model (Model~2) are applied 
to the spectrum of the inner 10$''$. The last column refers to the 
spectrum of the extended hot gas halo instead. Fluxes are given in units of 
$10^{-13}$ erg cm$^{-2}$ s$^{-1}$, luminosities in units of 
$10^{43}$ erg s$^{-1}$; both are corrected for absorption. 
$^*$ denotes a fixed parameter.}

\begin{center}
\begin{tabular}{llll}

\hline\hline 
  
& \bf{Model 1} & \bf{Model 2} & \bf{Hot Gas} \\
& Inner $10''$ & Inner $10''$ & \bf{Halo} \\
\hline

\bf{Apec(1)} & & & \\ 
$kT$ [keV] & $7.8^{+7.5}_{-2.8}$ & -- & -- \\
$f\,_{0.3 - 10\: keV}$ & $1.60^{+0.20}_{-0.17}$ & -- & -- \\   
\hline
     
\bf{Apec(2)} & & & \\ 
$kT$ [keV] & $0.80^{+0.04}_{-0.05}$ & $0.81\pm0.05$ & -- \\
$f\,_{0.3 - 10\: keV}$ & $0.37\pm0.06$ & $0.33^{+0.03}_{-0.05}$ &  -- \\   
\hline

\bf{Apec(3)} & & & \\
$kT$ [keV] & -- & -- & $0.38^{+0.05}_{-0.04}$ \\   
$f\,_{0.3 - 10\: keV}$ & -- & -- & $0.30\pm0.04$ \\   
\hline

\bf{Power-law(1)} & & & \\
$N_H$ [$10^{22}$cm$^{-2}$] & $41\pm6$ & $39\pm5$ & -- \\
$\Gamma$ & $1.9^*$ & $1.9^*$ & -- \\  
$f\,_{0.3 - 10 \:keV}$ & $53^{+12}_{-9}$ & $53^{+10}_{-9}$ & -- \\   
\hline
 
\bf{Power-law(2)} & & & \\
$N_H$ [$10^{20}$cm$^{-2}$] & -- & $6.8^{+2.3}_{-2.0}$ & -- \\
$\Gamma$ & -- & $1.9^*$ & -- \\  
$f\,_{0.3 - 10\: keV}$ & -- & $1.7\pm0.2$ & -- \\   
\hline
 
\bf{Fe K$\alpha$} & & & \\
$E$ & $6.34\pm0.04$ & $6.34\pm0.04$ & -- \\
$EW$ [eV] & $255\pm125$ & $242\pm120$ & -- \\     
$f\,_{0.3 - 10\: keV}$ & $0.72\pm0.35$ & $0.69\pm0.34$ & --  \\   
\hline

\bf{Total} & & & \\
$f\,_{0.3 - 10 \:keV}$ & 55.4 & 55.9 & 0.3 \\ 
$L\,_{0.3 - 10\: \mathrm{keV}}$ & 1.84 & 1.86 & 0.01 \\ 
\hline
  
$\chi^2/$dof & 65.0/75 & 62.0/75 & -- \\   
\hline

\label{chares}
\end{tabular}
\end{center}
\end{table}

\begin{figure}

\centerline{\epsfig{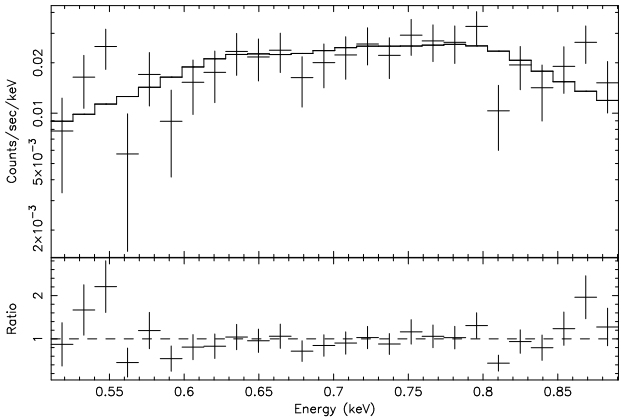}}
\caption{The $0.5-0.9$ keV unbinned spectrum of the extended hot gas halo. 
A single APEC model with a temperature of $0.38\pm0.04$ keV provides a 
satisfactory fit.}
\label{softxs}

\end{figure}

\subsection{The hot gas halo}

The ACIS-S spectrum of the extended hot gas halo, which is extracted from the 
region defined in Section 2.2, is source dominated only in the $0.5-0.9$ keV 
band. Given the limited amount of photons (315) detected 
in this band, we prefer to use Cash statistics on the unbinned spectrum. 
As displayed in Figure \ref{softxs}, the spectrum is consistent with a single 
temperature plasma component with $kT=0.38\pm0.04$ keV. We performed a Monte 
Carlo calculation to test the goodness of the fit and we found that 63\% of 
the 10$^4$ simulated spectra have a fit statistics less than that for the 
data. A fraction of approximately 50\% is expected, if the observed spectrum 
was produced by the model. If the $\chi^2$ statistics is applied to the 
spectrum grouped to have a minimum of 20 counts per bin, fixing the model 
parameters to those previously found (see Table~\ref{chares}), we obtain 
$\chi^2/dof=15/11$.

In this analysis we adopted a more conservative approach in subtracting 
the background in comparison to Paper~I. This resulted in a slightly smaller 
temperature for the extended hot gas halo with respect to the previous analysis, 
but in very good agreement with the XMM-{\em Newton} results. 

\section{Discussion}

A broad Fe K$\alpha$ emission line with $\sigma=0.26^{+0.37}_{-0.17}$ keV 
is detected at high significance in the broad band ($0.3-10$ keV) EPIC--PN 
spectrum ($\Delta\chi^2=11.4$ for 3 additional free parameters, corresponding 
to more than 99\% confidence level). The hypothesis of a broad iron line 
originating from the disc is plausible, but the high column density, obscuring 
the AGN up to about 4 keV, makes the detection of the redshifted wing of 
a relativistic iron line profile more difficult. The most likely explanation 
for such a broad line profile is the superposition of a narrow, unresolved, 
neutral iron line at 6.4 keV, probably associated with the AGN torus, and a 
blend of Fe XXV and Fe XXVI lines, associated with a hot plasma, in which this 
transitions are either thermally excited or induced by photoionization. 
Unfortunately, the modest statistics do not allow to singularly resolve the 
three possible lines. However the fluxes and equivalent widths measured by 
{\em Chandra} and XMM-{\em Newton} are consistent within the two observations. 

There is no evidence for extra absorbing material obscuring any of the thermal 
plasma components. This could be due to a non--spherical geometry of the gas 
surrounding the nuclear region, given the complex morphology of this merger, 
or it could be simply due to the low statistics. 
A slightly different column density is found to be obscuring the primary 
radiation between the two observations. Again this could be, somehow, 
associated to the complex morphology of the merger, since, for instance, the 
projected separation between the two nuclei is $\sim1''$, or it could be only 
an effect due to the low statistics available at high energies. 

According to the thermal emission model, the temperatures of the three 
plasma components found in Mrk~273 are remarkably similar to those observed in 
other ULIRGs (e.g. NGC~6240 and Arp~220) and in the local starburst galaxy 
NGC~253 \citep{Pie01}. The thermal emission at $\sim0.3$ keV, as clearly shown 
from the {\em Chandra} image, extends on a large scale ($\sim45$ kpc) and 
embraces the long tidal tail of the merger. Therefore, it must be associated 
with hot gas distributed in the halo. Interestingly, thermal emission at 
$\sim0.7$ keV seems to be ubiquitous in the spectra of ULIRGs, probably being 
associated with a nuclear or circumnuclear starburst \citep{Fra03}. The 
presence of a high temperature ($\gtrsim5$ keV) thermal component is less 
frequently observed, but not so unusual, being also detected in the 
XMM-{\em Newton} spectra of NGC~6240 \citep{Bol03} and Arp~220 \citep{Iwa05}. 
The ratio of the $2-10$ keV unabsorbed X--ray luminosity of the highest 
temperature thermal component to the bolometric luminosity in Mrk~273 is 
$L_{2-10\,\mathrm{keV}}/L_{FIR}\simeq7\times10^{-5}$ 
($L_{FIR}=L(8-1000\,\mu\mathrm{m}\sim L_{bol}$), a factor of 20 smaller than 
in NGC~6240 and a factor of 5 larger than in Arp~220. 

On the other hand, the second model investigated suggests that a 
secondary hard X--ray component originates from the reflection of the primary 
continuum on some photoionized circumnuclear material. This is often 
observed in Compton--thick Seyfert 2 galaxies \citep{Sam01,Kin02,Don03,Bia05}, 
which, in some cases, also show evidence for starburst activity 
(e.g. Circinus, Ruiz et al. 2001; NGC~1068, Smith \& Wilson 2003). As was 
observed in Paper I, two bright clumps of soft X--ray emission coincide with 
[OIII] emission. This piece of evidence seems to support the hypothesis of a 
photoionized gas surrounding the nucleus, since in several Compton--thick 
Seyfert galaxies, the extended soft X--ray emission expected to be produced by 
reflection from that photoionized gas has been observed to coincide with the 
[OIII] emission and the ionization cone (e.g. Sako et al. 2000; 
Young et al. 2001). 

The absorption corrected hard X--ray luminosity of Mrk~273 is 
$L_{2-10\,\mathrm{keV}}\sim7\times10^{42}$ erg s$^{-1}$, which corresponds to 
$\sim2\times10^{-3}$ of the far--IR luminosity ($L_{FIR}\sim10^{12}L_{\odot}$, 
e.g. Genzel et al. 1998). Such a ratio approaches values of pure starbursts 
and is orders of magnitude smaller than in Compton-thin AGNs \citep{Ris00}. 
According to the thermal emission model, the absorption corrected $0.3-10$ keV 
luminosity due to the three thermal components would be approximately 
$0.7\times10^{42}$ erg s$^{-1}$. Comparable soft X--ray luminosities have been 
measured for NGC~6240 \citep{Net05} and other ULIRGs in which the starburst 
contribution to the bolometric luminosity is dominating \citep{Fra03}.
From the photoionized gas model the resulting thermal contribution to the 
X--ray luminosity is about a factor of 3 smaller. 
For Mrk 273 the unobscured AGN X-ray luminosity is a more modest fraction of 
the bolometric luminosity than in typical AGN. This suggests additional strong 
star formation which will naturally explain the thermal X-ray components 
invoked in one of our two scenarios. The presence of strong star formation 
agrees with the detection from mid-infrared spectroscopy \citep{Gen98} of both 
very powerful star formation and an AGN in Mrk 273. 

It must be mentioned that, besides the two models discussed above, 
other interpretations are possible. In fact X--ray emission from luminous 
radio supernovae like SN1995J \citep{Fox00}, or the integrated contribution 
from low-- and high--mass X--ray binaries, as observed from the spectra of 
local starburst galaxies \citep{Per02}, may both give rise to a very hot 
($kT\sim5$ keV) thermal component, or a $\Gamma\sim2$ power law. Following 
\citet{Ken98}, the star formation rate estimated assuming that the entire 
$L_{FIR}$ is due to starburst would be $\sim200$ $M_{\odot}$ yr$^{-1}$. 
Therefore, the inferred supernova rate is $\sim2$ yr$^{-1}$, which is in 
agreement with the value estimated from radio observations \citet{Bon05}. 
On the other hand, given a 200 $M_{\odot}$ yr$^{-1}$ star formation rate, 
the estimated emission from X--ray binaries, following \citet{Fra03}, would be 
$L_{2-10\,\mathrm{keV}}\sim2\times10^{41}$ erg s$^{-1}$, which is very close 
to the observed value ($\sim3\times10^{41}$ erg s$^{-1}$). 
Finally, it is worth noting that X--ray spectra from supernovae have shown 
evidence for strong Fe K line at 6.7 keV (e.g. SN1986J, Houck et al. 1998), 
therefore could also account for this feature, while emission from X--ray 
binaries can not.

\section{Conclusions}

We analyzed the X--ray spectral properties of Mrk~273 combining the high 
throughput of XMM-{\em Newton} with the high spatial resolution of 
{\em Chandra}. 

From the XMM-{\em Newton} spectrum we found a broad Fe K$\alpha$ emission line 
with a high statistical significance ($\gtrsim$99\% c.l.). We suggested the 
superposition of multiple unresolved iron line features: one from neutral iron 
at 6.4 keV ($EW\sim170$ eV) and one from a blend of Fe XXV at 6.7 keV 
($EW\sim120$ eV) and Fe XXVI at 6.97 keV ($EW<85$ eV).

We tested two different models capable of reproducing the possible ionized 
iron lines detected: one through thermal emission, therefore involving the 
presence of a hot gas associated with a starburst, and the other one through 
reflection of the primary continuum by some optically thin photoionized gas. 
Unfortunately, given the available statistics, we can not rule out any of the 
two possible explanations. 
Some indications that a reflection componet is indeed present in the nuclear 
region of Mrk~273 may come from polarization measures, which are already 
available in the IR \citep{Sie01}, but still not in the X--rays. 
High resolution spectroscopy could be helpful in order to detect some of 
the forbidden lines or the Radiative Recombination Continua (RRC), which are 
considered as typical signatures of photoionization \citep{Por00,Lie99,Lie96}. 

\appendix

\section{The unabsorbed Seyfert~2 Mrk~273x}
\label{app:m273x}

Over recent years, much attention has been directed to the study of unabsorbed 
Seyfert~2 galaxies (e.g. Ptak et al. 1996; Panessa \& Bassani 2002; 
Gallo et al. 2005). These galaxies are of interest for two reason: they may 
account for as much as 30\% of the Seyfert~2 population \citep{Pan02}, and 
they are a challenge to explain with the existing Seyfert unification model.

Mrk~273x ($z =$ 0.458) is one of the better known unabsorbed Seyfert~2 
galaxies, perhaps because it falls in the field-of-view of most observations
of Mrk~273. In this appendix, we would like to discuss the properties of 
Mrk~273x as seen during the XMM-{\em Newton} observation of Mrk 273. 

In the EPIC-PN image, Mrk~273x falls directly in a chip gap; hence we used the 
combined MOS 1/2 data for spectral analysis. The combined spectrum is source 
dominated in the $0.3-10$~keV band ($0.44-14.6$~keV in the rest-frame).
The total number of source plus background counts is 526, whereas the total
number of background counts is 65 (scaled to source cell size).

\begin{figure}
\centerline{\epsfig{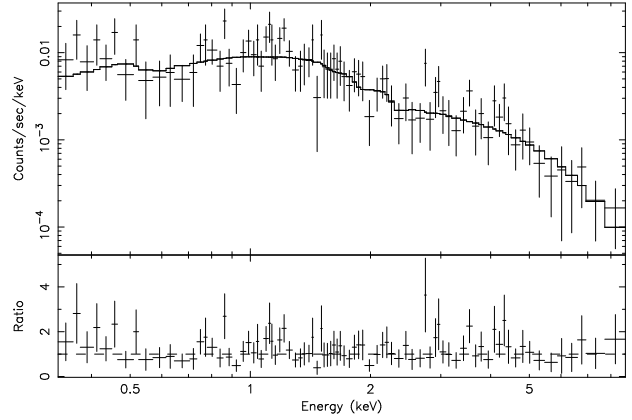}}
\caption{The combined MOS 1 and MOS2 $0.3-10$~keV data fitted with a power law
modified by Galactic absorption.}
\label{fig:mrk273xpo}
\end{figure}

The best-fit continuum model of Paper I to the $Chandra$ data of Mrk~273x was 
a power law plus a weak thermal component. Significantly detected above the 
continuum was an emission line at 1.82~keV, which the authors suggested could 
be due to Si~XVI. The XMM-{\em Newton} data are consistent with only a 
power law ($\Gamma = 1.49 \pm 0.12$) modified by Galactic absorption 
($\chi^2/dof=60.2/82$; Figure~\ref{fig:mrk273xpo}). 

The 90\% confidence upper limit on the intrinsic absorption is 
$<4.5\times10^{20}$ cm$^{-2}$, at least a factor of three less than what was 
required in the $Chandra$ model. 

Multi-component continuum models (e.g. blackbody plus power law, double 
power law, or a broken power law) also fit the data well, but they are not
significant improvements over the single power law. Neither the addition of a 
$\sim1.82$~keV emission feature is required by the data ($\Delta\chi^2=1$ for
2 additional free parameters; $EW<33$~eV). Positive residuals at about 4.3~keV 
can be modelled with a narrow Gaussian profile with a rest-frame energy of 
$E=6.29^{+0.14}_{-0.34}$~keV. The addition of the Gaussian profile is an 
improvement to the residuals, but it is not statistically significant 
($\Delta\chi^2=3$ for 3 additional free parameters).

Based on the single power law fit, the unabsorbed $0.5-10$~keV flux is
($1.43^{+0.22}_{-0.18})\times10^{-13}$ erg s$^{-1}$ cm$^{-2}$
($1.18\times10^{-13}$ erg s$^{-1}$ cm$^{-2}$ in the $2-10$~keV band). 
The intrinsic $0.3-10$~keV unabsorbed luminosity is $9.2\times10^{43}$ erg 
s$^{-1}$. The broad band flux measured during this XMM-{\em Newton} 
observation is comparable to the flux observed during the $Chandra$ 
observation two years earlier. The $\sim22$~ks light curve of Mrk~273x shows 
no variability and is perfectly fitted by a constant.

\acknowledgements

We thank G\"unther Hasinger and Andrea Comastri for useful discussions, 
Stefano Bianchi and Paolo Tozzi for their help on the reduction and spectral 
analysis of the XMM-{\em Newton} and {\em Chandra} data and the anonymous 
referee for valuable comments and suggestions. This paper is based on 
observations obtained with XMM-{\em Newton}, an ESA science mission with 
instruments and contributions directly founded by ESA Member States and NASA.

\end{document}